\documentclass{nature}

\usepackage[top=0.85in,left=1.55in,right=1.5in,footskip=0.75in]{geometry}

\usepackage{amsmath,amssymb}
\usepackage{nameref,hyperref}
\usepackage{mathptmx, siunitx, adforn}
\usepackage{graphicx}
\usepackage[all]{nowidow}

\usepackage{textcomp,marvosym}

\usepackage[right]{lineno}

\spacing{1.2}

\newcommand*\patchAmsMathEnvironmentForLineno[1]{
  \expandafter\let\csname old#1\expandafter\endcsname\csname #1\endcsname
  \expandafter\let\csname oldend#1\expandafter\endcsname\csname end#1\endcsname
  \renewenvironment{#1}
     {\linenomath\csname old#1\endcsname}
     {\csname oldend#1\endcsname\endlinenomath}}
\newcommand*\patchBothAmsMathEnvironmentsForLineno[1]{
  \patchAmsMathEnvironmentForLineno{#1}
  \patchAmsMathEnvironmentForLineno{#1*}}
\AtBeginDocument{
\patchBothAmsMathEnvironmentsForLineno{equation}
\patchBothAmsMathEnvironmentsForLineno{align}
\patchBothAmsMathEnvironmentsForLineno{flalign}
\patchBothAmsMathEnvironmentsForLineno{alignat}
\patchBothAmsMathEnvironmentsForLineno{gather}
\patchBothAmsMathEnvironmentsForLineno{multline}
}

\usepackage{microtype}
\DisableLigatures[f]{encoding = *, family = * }

\usepackage{array}

\newcolumntype{+}{!{\vrule width 2pt}}

\newlength\savedwidth

\newcommand\thickhline{\noalign{\global\savedwidth\arrayrulewidth\global\arrayrulewidth 2pt}%
\hline
\noalign{\global\arrayrulewidth\savedwidth}}

\title{Surround suppression explained by long-range recruitment of local competition, in a columnar V1 model} 

\author{Hongzhi You\textsuperscript{1,2},
Giacomo Indiveri\textsuperscript{1,*},
Dylan R. Muir\textsuperscript{3,*}}

\begin{document}

\maketitle

\begin{affiliations}
 \item Institute of Neuroinformatics, University / ETH Z\"{u}rich, Z\"{u}rich, Switzerland
 \item Key Laboratory for NeuroInformation of Ministry of Education, Center for Information in BioMedicine, University of Electronic Science and Technology of China, Chengdu, China
 \item Biozentrum, University of Basel, Basel, Switzerland
\end{affiliations}

\begin{abstract}
  Although neurons in columns of visual cortex of adult carnivores and primates share similar orientation tuning preferences, responses of nearby neurons are surprisingly sparse and temporally uncorrelated, especially in response to complex visual scenes. The mechanisms underlying this counter-intuitive combination of response properties are still unknown. Here we present a computational model of columnar visual cortex which explains experimentally observed integration of complex features across the visual field, and which is consistent with anatomical and physiological profiles of cortical excitation and inhibition. In this model, sparse local excitatory connections within columns, coupled with strong unspecific local inhibition and functionally-specific long-range excitatory connections across columns, give rise to competitive dynamics that reproduce experimental observations. Our results explain surround modulation of responses to simple and complex visual stimuli, including reduced correlation of nearby excitatory neurons, increased excitatory response selectivity, increased inhibitory selectivity, and complex orientation-tuning of surround modulation.

\end{abstract}

In species with highly developed neocortices, such as cats and primates, cortical neurons are grouped into columns that share functional similarities\cite{Mountcastle_etal55}. In primary visual cortex, columns of neurons have highly similar preferred orientations of visual stimuli\cite{Hubel_Wiesel62, Hubel_Wiesel68}. However, given that neurons in a column share the same retinotopic location and have common orientation preferences, their firing activity is surprisingly poorly correlated\cite{Yen2007, Ecker_etal10, Martin_Schroder13}, even in response to drifting grating stimuli\cite{Ecker_etal10, Martin_Schroder13}.

Since natural sensory inputs are highly temporally correlated\cite{Kersten87, Schwartz_Simoncelli01}, an active mechanism is required to reduce correlations, and consequently, to improve information coding efficiency\cite{Schwartz_Simoncelli01, Wiechert_etal10}. This is beneficial because strong correlations across a neuronal population can impair the ability to extract information from their response to sensory stimuli\cite{Shamir_Sompolinsky04, Averbeck_etal06}. ``Sparse coding'' of responses to sensory stimuli is therefore a valuable goal for cortex: sparse coding serves to increase storage capacity\cite{willshaw1969non, olshausen2004sparse} and information efficiency\cite{Shamir_Sompolinsky04, Averbeck_etal06} of cortical populations. In visual cortex, the functional relationships between nearby neurons is modulated by the information from the visual surround: wide-field stimulation with natural scenes promotes more selective and less correlated excitatory activity\cite{Vinje_Gallant00, Haider_etal10}, while inhibitory activity becomes stronger and less selective\cite{Haider_etal10}.

How does this reduction in response correlation come about, given the prevalence of strong spatial and temporal correlations present in natural visual scenes\cite{Kersten87, Schwartz_Simoncelli01}, and given that neurons in a column share common preferences for visual features? Several neural models have been proposed to reduce correlations in network activity, including non-linearity of spike generation, synaptic-transmission variability and failure, short-term synaptic depression, heterogeneity in network connectivity, intrinsic neuronal properties and recurrent network dynamics \cite{de2007correlation, hertz2010cross, rosenbaum2013short, bernacchia2013decorrelation, Helias2014}. A particularly appealing form of recurrent dynamics is that involving inhibitory feedback loops, which are abundant in cortical networks\cite{Schwartz_Simoncelli01, renart2010asynchronous, Wiechert_etal10, Ecker_etal10, Tetzlaff2012, Helias2014, pernice2011structure}. When configured appropriately, inhibitory feedback promotes competition between the activity of a set of excitatory neurons, such that weaker responses are suppressed in a non-linear fashion\cite{Coultrip_etal92, Douglas_etal94, Douglas_Martin07, Rutishauser_Douglas09}.

As discussed above, competition directly reduces correlations within a network (in the sense of making correlation coefficients more negative). Local competition within a cortical column would result in a few ``winning'' neurons with increased activity, while a majority of ``losing'' neurons would decrease their activity. Grating stimuli presented in the visual surround predominately suppress neural responses\cite{blakemore1972lateral, Nelson_Frost78, Bonds89, Nelson91, DeAngelis_etal92, Walker_etal99, Freeman_etal02, Shushruth2012}, with more than 50\% of neurons reducing their firing rate\cite{Nelson91}. A comparatively smaller proportion of neurons undergo facilitation in response to surround stimulation\cite{Sillito_etal95, Bonds89, Nelson91}. We therefore suggest that the physiology of surround suppression and facilitation in columnar cortex is consistent with local competitive mechanisms operating within a cortical column.

Local excitatory connections are sparse in cortex, with maximum connection probabilities between closest proximal cells (i.e. $\approx\SI{50}{\micro\meter}$) of only \SIrange{20}{30}{\%}\cite{Yoshimura_etal05, Lefort_etal09, Ko_etal11, Ishikawa_etal14} and with connection probability falling off sharply with distance \cite{matsuzaki2008three, boucsein2011beyond}. In contrast, local connections between excitatory and inhibitory neurons are dense in rodents\cite{Martin11, Fino_Yuste11, Bock_etal11, Hofer_etal11}. In animals with columnar visual cortices, inhibitory inputs are simply integrated from the nearby surrounding tissue\cite{Marino_etal05}, suggesting a similar pattern of dense local connectivity. Long-range excitatory connections (i.e.~\SIrange{500}{1500}{\micro\meter}) within columnar visual cortex are made selectively between points across the cortical surface with similar functional preferences\cite{Gilbert_Wiesel89, Malach_etal93, Yoshioka_etal96, Bosking_etal97, Muir_etal11, Muir_etal11a, Martin_etal14}. Similarly, excitatory connections in rodents are made selectively\cite{Yoshimura_Callaway05, Yoshimura_etal05}, between neurons with correlated functional properties\cite{Ko_etal11, Hofer_etal11, Cossell_etal15}.

Here we propose a computational model that, consistent with anatomy,  exploits both long-range and local excitatory interaction between cortical circuits connecting neurons in the superficial layers of cortex to explain the observed sparse response properties of cortical neurons. Local excitatory interactions result in local competition within a cortical column, which is recruited and modulated by information conveyed over long-range excitatory projections, including from the visual surround. The network proposed models the superficial layers of cat primary visual cortex (area 17), including several populations designed to simulate distinct portions of the visual field (``centre'' and ``surround''). We present the model's response to a range of simulated visual stimuli, designed in analogy to experimental investigations of centre/surround visual interactions, and show how the mechanism of local competition is recruited by visual stimulation to reduce local correlations and to suppress neuronal responses. The mechanism of within-column competition explains the complex physiology of suppressive and facilitatory influences from the visual surround \cite{Nelson_Frost78, Walker_etal99, Nelson91, Shushruth2012, Hashemi_Lyon12}.

\section*{Results}
\subsection{Network architecture.}

\begin{figure}
\centering
\includegraphics[angle = 0, width = 4.0 in]{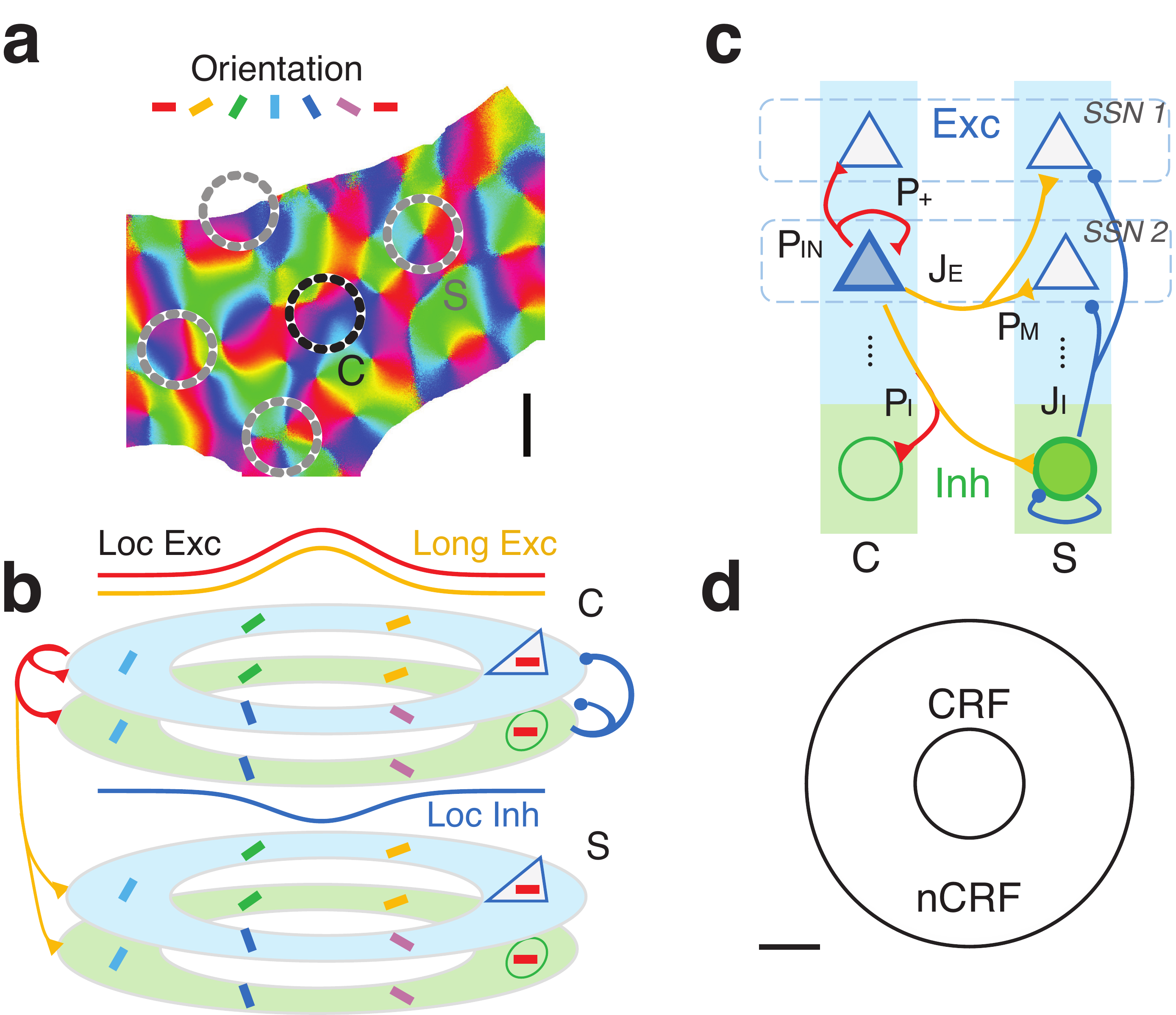}
\caption[Network architecture.]{\textbf{The network architecture of the center-surround model.} \textbf{(a)}~We simulated several populations representing non-overlapping locations in primary visual cortex (dashed cicrles), covering distinct locations of the visual field.  Each population was simulated as a ring of neurons considered to span an orientation hypercolumn, i.e. containing a set of columns with a complete ordered sequence of orientation preferences. One population~(``C'') corresponded to the centre of visual stimulation; the others corresponded to the visual surround~(``S''). Scale bar:~\SI{1}{mm} \textbf{(b)}~Excitatory (triangles) and inhibitory neurons (circles) in each population were arranged in orientation columns around the ring, with preferred orientations indicated by coloured bars. Local excitatory and inhibitory connections within each population, as well as long-range excitatory connections between populations, were modelled as Gaussian fields over difference in preferred orientation (curves in~b). For simplicity only projections from a single population are shown (``C'' ring; upper); connections were made identically within and between each population in the model. \textbf{(c)}~Connections from single neurons were made both within a column, and between populations. Excitatory neurons within a column were distributed evenly across several subnetworks (SSNs; see Methods for details of parameters). A proportion of local excitatory synapses was reserved to be made only with other neurons within the same SSN~($P_+$). Long-range excitatory connections were also sensitive to SSN membership, under the parameter~$P_M$. $J_E$,~$J_I$:~Strength of excitatory~($E$) and inhibitory~($I$) synapses. $P_{IN}$,~$P_I$:~Fraction of synapses onto excitatory~($IN$) and inhibitory~($I$) targets made locally within the same hypercolumn, as opposed to long-range projections to other hypercolumns. For simplicity, only connections from a single excitatory and inhibitory neuron are shown; projection rules are identical for all neurons in the model. \textbf{(d)} Placed in visual space, the central population corresponded to approximately~\ang{1} of visual space; the surround populations were defined to cover approximately~\ang{4.5} of visual space. Scale bar:~\ang{1}v}
\label{Fig1}
\end{figure}

We developed a spiking network model for adult columnar primary visual cortex, composed of~\SI{7} populations of neurons, each representing a distinct location on the visual field (Fig.~\ref{Fig1}a; see Methods). Each population consisted of a ring representing one hypercolumn --- a full ordered sequence of preferred orientations corresponding to approximately~\SI{1}{mm} of cat area~17~(V1)\cite{Muir_etal11a}. One population was arbitrarily chosen as the \textit{center} of visual stimulation; the other populations represented \textit{surround}ing areas of visual space (i.e. the visual surround). Each population was composed of a series of columns consisting of excitatory and inhibitory neurons, where each column contained neurons with a common preferred orientation. Inhibitory neurons made only short-range recurrent connections within their source population; in contrast, excitatory neurons made wider-ranging recurrent connections within their source population, as well as long-range connections to the other populations in the model (Fig.~\ref{Fig1}b). All connections were made symmetrically following Gaussian profiles over difference in preferred orientation, taking into account the ring topology within each population. Long-range connections were therefore biased to connect columns with similar orientation preference, as is observed in cat visual cortex\cite{Bosking_etal97, Muir_etal11a}. In this paper, we refer to this form of functional synaptic specificity as~\textit{$\theta$-specific} (i.e. orientation-specific).

Within each column, excitatory neurons were assumed to form sub-populations (``specific subnetworks'', or SSNs; Fig.~\ref{Fig1}c), which have a higher-than-chance probability of forming recurrent excitatory connections. In contrast, inhibitory neurons have equal probabilities of forming recurrent inhibitory connections to all neurons within a column, regardless of SSN membership. This architecture is known to exist in rodent visual cortex\cite{Yoshimura_Callaway05, Yoshimura_etal05, Ko_etal11, Hofer_etal11, Cossell_etal15}, but has not been examined experimentally or computationally in columnar visual cortex. In this paper we refer to this form of specificity as \textit{SSN-specific} connectivity. SSN-specificity in our model could be replaced by the presence of strong recurrent excitatory feedback on the scale of single cortical neurons, if necessary. The strength of SSN-specificity was under the control of a parameter $P_+$ (see Methods). As $P_+\rightarrow1$, all excitatory connections are made within the same SSN. As $P_+{\rightarrow}1/M$, all synapses are distributed equally across all SSNs; equivalent to fully random connectivity within a column (here, $M$ is the number of SSNs in each column). Long-range connections were also SSN-specific under the control of a parameter $P_M$ (with analogous definition as $P_+$). All parameter values were selected to be realistic estimates for cat area 17 (see Table~\ref{Table1}).

Our model examined modulation of orientation-tuned responses, caused by inputs from the visual surround, carried by long-range excitatory connections within the superficial layers of columnar cortex. Our model did not investigate the emergence of orientation tuning, which occurs from convergence of thalamic afferents into cortex\cite{Jin_etal11}. We assumed that the neurons in our model resided in the superficial layers of cortex, and therefore received orientation-tuned input primarily from layer 4.

\subsection{Neurons within a column are similarly tuned, but without temporally correlated responses.}

\begin{figure}
\centering
\includegraphics[angle = 0, width = 5.0 in]{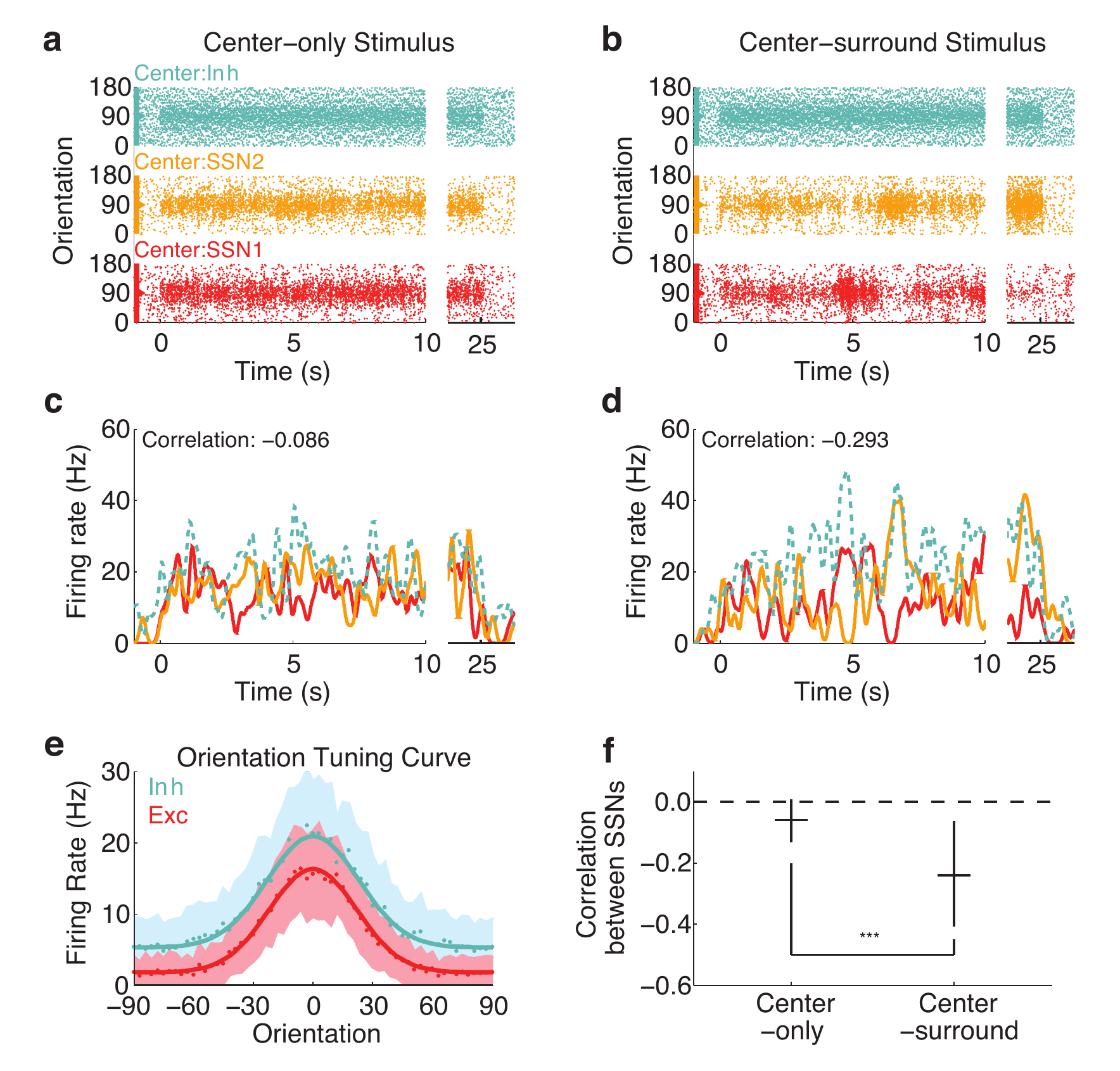}
\caption[Reduced correlation in response to grating stimuli]{\textbf{Neurons within a column are not temporally correlated in response to centre-surround grating stimulation.} \textbf{(a--b)}~Spiking responses of neurons in the centre population, in response to centre-only~(a) and centre-surround stimulation~(b) with \ang{90} orientation stimulus. Inhibitory neurons: blue-green; two of four excitatory SSNs: red and yellow. \textbf{(c--d)}~Firing rates over time for neurons with~\ang{90} orientation preference (colours as in a--b). \textbf{e}~Orientation tuning curves for excitatory and inhibitory neurons are similar. Under centre-only stimulation (a and~c), excitatory neurons within a column respond together, since they share a common preferred orientation, but are temporally decorrelated (correlation coefficient close to zero; c and~\textbf{f}). Under wide-field stimulation (b and~d), responses of neurons within the same column become more negatively correlated (negative correlation coefficient; d and~f). ***~$p<0.001$.}
\label{Fig2}
\end{figure}

We tested the response of our model to simulated grating visual stimuli, presented first to the classical receptive field only (cRF; centre-only stimulus), and under wide-field stimulation (centre-surround stimulus) (Fig.~\ref{Fig2}). Stimulation of the cRF with grating stimuli of a single orientation provoked a response over the central population according to the similarity between the stimulus and the preferred orientation of each column (Fig.~\ref{Fig2}a).

Orientation tuning curves were similar between excitatory and inhibitory neurons (Fig.~\ref{Fig2}e; tuning widths \ang{25.9} half-width at half-height for excitatory neurons and \ang{27.1} for inhibitory neurons; $P<10^{-10}$, t-test, 200~neurons). Although neurons within the same column were tuned to identical preferred orientations and responded with a higher average firing rate to the same stimuli, temporal patterns of activation were weakly but significantly negatively correlated on average within a column (Fig.~\ref{Fig2}c and~f; median corr. $-0.06$; $P<10^{-10}$, rank-sum test, 200~trials).

Wide-field presentation of simulated grating stimuli provoked stronger negative correlations between neurons within a column (Fig.~\ref{Fig2}b, d and~f; median corr.~$-0.06$ vs~$-0.24$; $P<10^{-10}$, rank-sum test, 200~trials). Wide-field stimulation also increased the firing rate of the inhibitory population (\SIrange{21.2}{26.0}{Hz}; $P<10^{-10}$, t-test, 200~trials) and decreased the firing rate of the excitatory population (\SIrange{16.4}{11.7}{Hz}; $P<10^{-10}$, t-test, 200~trials), consistent with experimental observations in visual cortex\cite{Haider_etal10}. The orientation tuning width of inhibitory neurons increased slightly under wide-field stimulation (\SIrange{27.1}{28.0}{\degree}; $P<10^{-10}$, t-test, 200~trials).

\subsection{Non-random excitatory connectivity promotes negative correlation of neural responses.}

\begin{figure}
\centering
\includegraphics[angle = 0, width = 4 in]{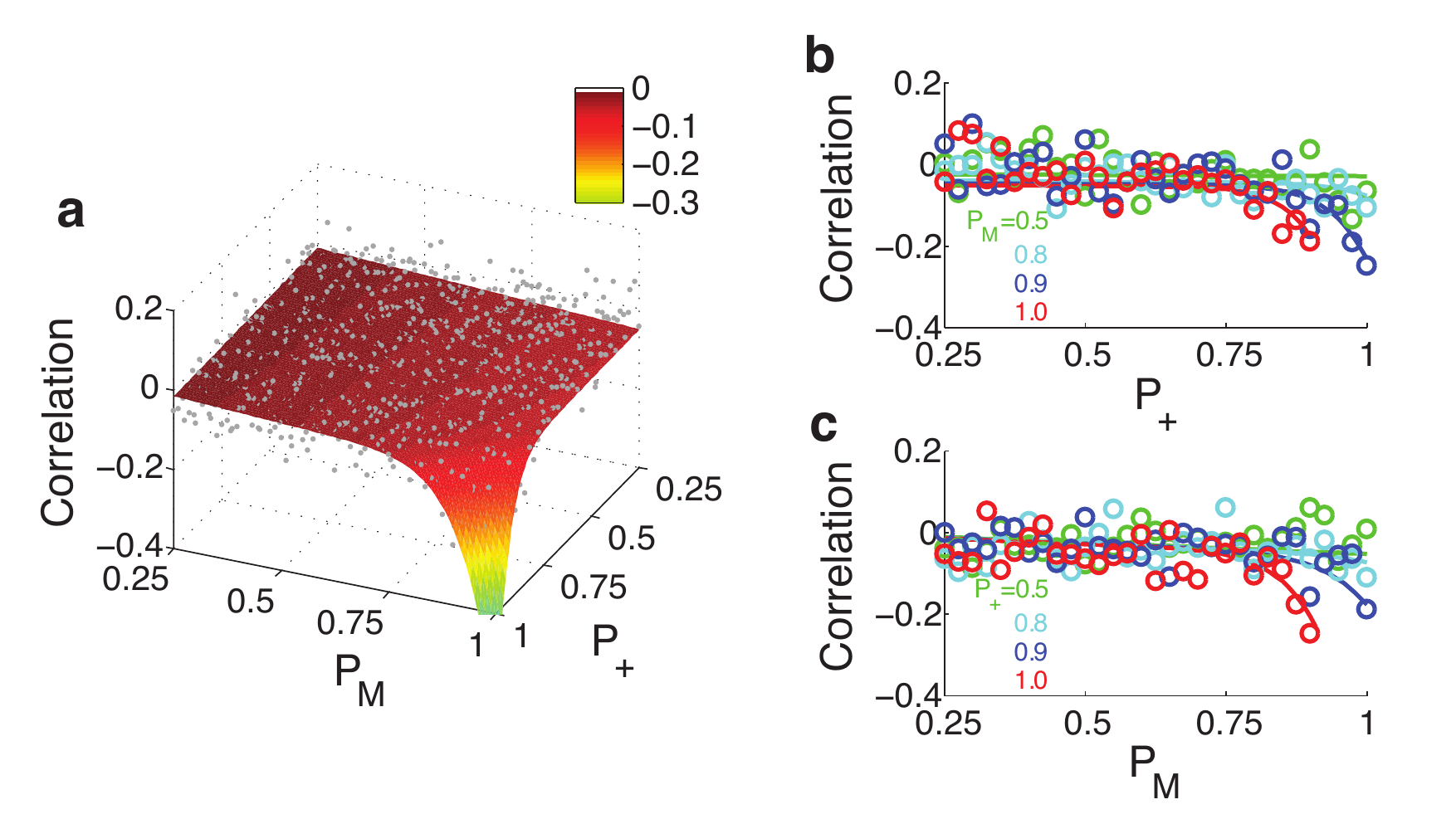}
\caption[Numerical analysis]{\textbf{Non-random excitatory connectivity underlies competition and reduced correlation in response to centre-surround stimulation.} \textbf{(a--c)} Correlation between the neurons in a column, as a function of the SSN-specificity parameters $P_+$ and $P_M$ (see Methods; Fig.~\ref{Fig1}). Grey dots and circles: measurements from individual simulations of the spiking network model. Surface in (a) and curves in (b--c): smooth fit to individual simulations. Both local recurrent specificity ($P_+$) and long-range specificity ($P_M$) promote competition within single columns (negative correlation coefficients).}
\label{Fig3}
\end{figure}

What parameters of cortical connectivity lead to competition in response to centre-surround stimulation? We explored the dependence of competition on the degree of non-random connectivity, both local ($P_+$) and long-range projections ($P_M$; see Methods). We simulated the presentation of wide-field stimulation with grating stimuli, as in Fig.~\ref{Fig2}, and measured the average correlation coefficient between neurons in the same column. Measurements of correlation coefficients over many network instances with varying $P_+$ and $P_M$ are shown in Fig.~\ref{Fig3}. In all cases, connections within and between populations were $\theta$-specific. However, competition depended strongly on non-random excitatory connectivity, such that when connections were made without local or long-range SSN-specificity (i.e. low $P_+$ and $P_M$), responses within a column were correlated. In contrast, when connections were highly non-random (i.e. $P_+$, $P_{M}{\rightarrow}1$) then responses within a column were negatively correlated.

The effects of local and long-range non-random connectivity are mutually supportive. When either of local or long-range connections are made SSN-specific then weak competition is introduced. However, when both connection pathways are SSN-specific then competition is significantly strengthened.

We hypothesised that the negative correlations introduced by non-random connectivity depends on the mechanism of competition within a cortical column. We therefore used nonlinear dynamical analysis to explore the presence of competition within a mean-field version of the model, and the dependence of competition on network parameters. Our mean-field model included only two populations from the full spiking model, with each population reduced to a single column; that is, the orientation-selective profile of connectivity is neglected (see details in \textbf{Methods}). In this analysis we systematically vary the parameters of the model under centre-only and centre-surround stimulation, and characterise the strength of competition within a column (Competition Index --- CI; see Methods).

\begin{figure}
\centering
\includegraphics[angle = 0, width = 4 in]{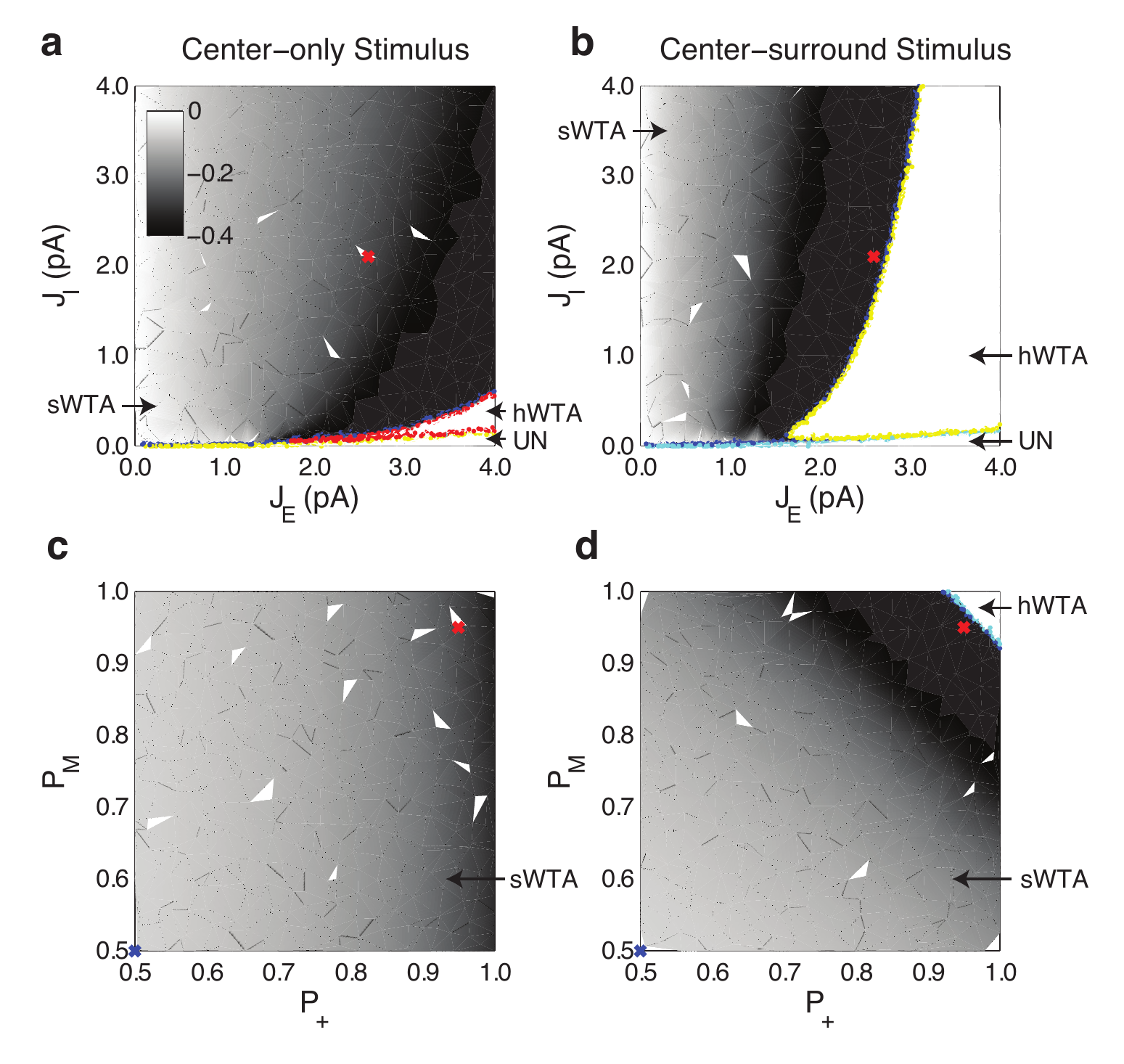}
\caption[Theoretical analysis]{\textbf{Dependence of competition on stimulation type and model parameters.} Both synaptic strength (\textbf{a--b}; inhibitory: $J_I$; excitatory: $J_E$) and degree of SSN-specificity (\textbf{c--d}; local: $P_+$; long-range: $P_M$) affect the strength of competition within a column (negative competition index; grey shading. See Methods). Centre-surround stimulation~(b and~d) generally increases the strength of competition. Both soft and hard competitive regimes exist. sWTA: soft winner-take-all (WTA) regime; hWTA: hard WTA regime; UN: unbalanced regime.}
\label{Fig4}
\end{figure}

Results of this analysis are shown in Fig.~\ref{Fig4}. In general, increasing the strength of excitatory connections ($J_E$) increased the strength of competition, and the opposite was true for the strength of inhibitory connections ($J_I$). Although competition is mediated via disynaptic inhibitory interactions between excitatory neurons, competition also requires strong excitatory interaction within SSNs, and increasing the strength of inhibition reduces the ability of excitatory neurons to recruit others within the same SSN.

Under our estimates for synaptic strengths approximating cortical connectivity (red crosses in Fig.~\ref{Fig4}a and~b; Table~\ref{Table1}; see Methods), activity within a column is balanced. Neurons within a column undergo mutual soft winner-take-all competitive interactions (sWTA; gray shading in Fig.~\ref{Fig4}; \cite{Douglas_Martin07}). In this regime, increasing the activity of one excitatory neuron results in a decrease of the activity of the other neurons within the column, but is not able to reduce their activity to below the firing threshold. If excitation is strengthened, a regime of hard competition is reached (hWTA), whereby only a single excitatory neuron can be active at a given time. If inhibition is made too weak, then activity within a column becomes unbalanced and saturates, and no competition is possible (UN).

Providing wide-field input to center and surround modules strengthens and changes the profile of competition within a column. For a given choice of parameters, competitive interactions are strengthened by surround stimulation (compare CI at red crosses in Fig.~\ref{Fig4}a and~b; c~and~d). This occurs simultaneously with a shifting of the parameter regions of the model, such that the size of the region of hard competitive interactions is increased (Fig.~\ref{Fig4}b). Consistent with competition being responsible for reduced correlation in the spiking model, the strength of competition was related to the degree of local and long-range non-random connectivity ($P_+$ and $P_M$; Fig.~\ref{Fig4}c and~d).

\subsection{Local competition coupled with tuned long-range connections explains orientation-tuned surround suppression.}

\begin{figure}
\centering
\includegraphics[angle = 0, width = 4.0 in]{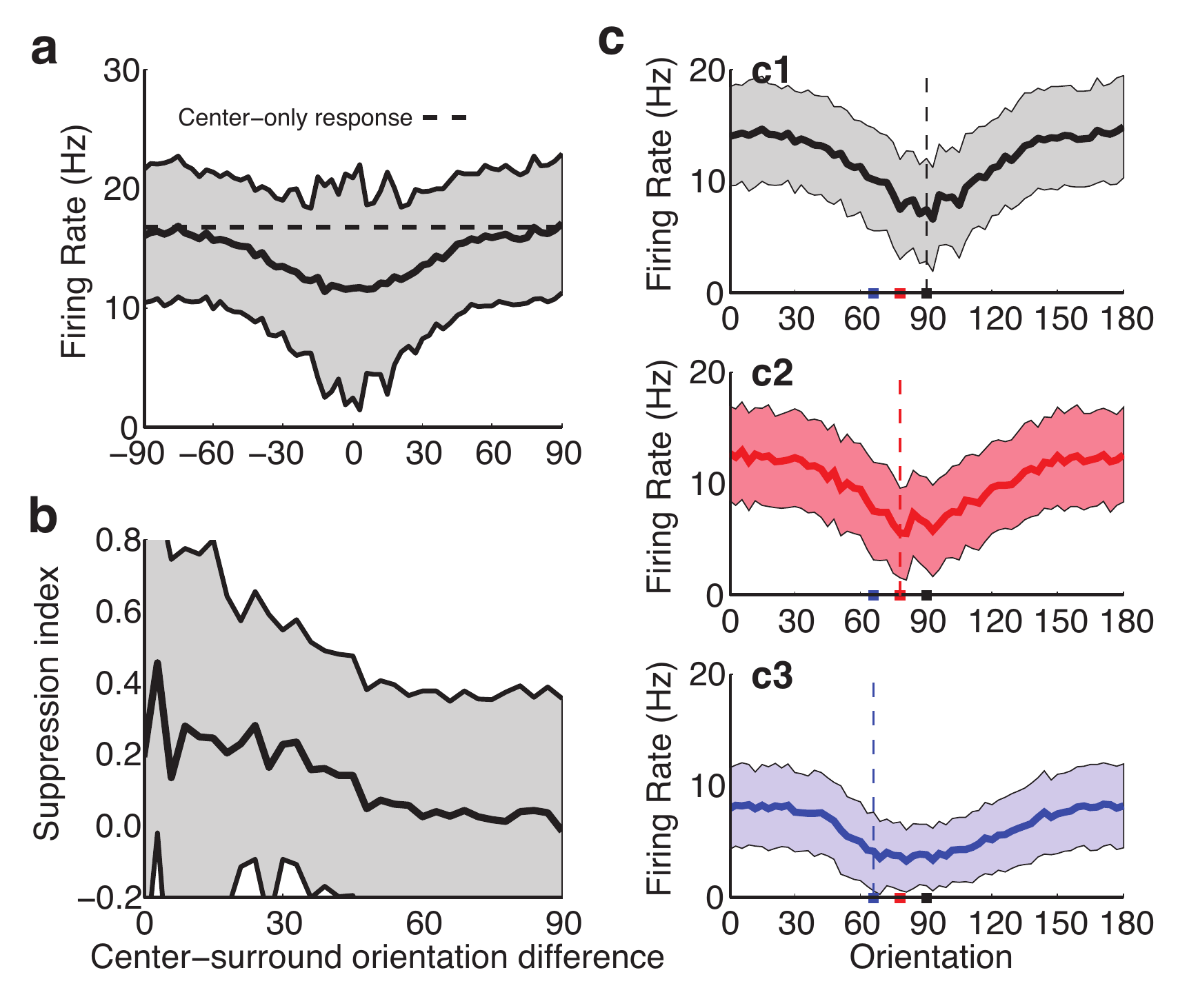}
\caption[Orientation-tuned suppression.]{\textbf{Orientation-tuned suppression under centre-surround stimulation.} \textbf{(a)} Centre-surround grating stimulation (black line: mean response; shading: std. dev.) provokes suppression on average, compared with centre-only stimulation (dashed line: response for centre-only stimulation at preferred orientation). \textbf{(b)} In agreement with experimental findings, our model exhibits orientation-tuned suppression that is strongest when centre and surround orientations are aligned\cite{Nelson_Frost78, Walker_etal99, Nelson91}. \textbf{(c)} Responses of neurons preferring vertical gratings ($90^\circ$) that exhibit suppression under center-surround stimulation with gratings (orientation of surround stimulus indicated on x-axis). Mean response (thick lines) and standard deviations (shading) are shown. The profile of suppression shifts depending on the orientation presented to the cRF (colours of curves corresponding to pips at bottom, indicating orientation of cRF stimulus). The orientation that provokes maximum surround suppression depends on the orientation presented to the classical receptive field of a neuron, not on the preferred orientation of that neuron\cite{Shushruth2012}.
}
\label{Fig5}
\end{figure}

In visual cortex, the strength of suppression induced by wide-field grating stimulation depends on the relative orientations of the grating stimuli presented in the cRF and in the visual surround\cite{Nelson_Frost78, Walker_etal99, Nelson91, Shushruth2012}. We examined the tuning of suppression in our model by simulating the presentation of two gratings to the centre and surround populations, while varying the relative stimulus orientations (Fig.~\ref{Fig5}). Consistent with experimental findings, the strongest suppression occurred in our model when the orientations of the center and surround stimuli were aligned ($\Delta\theta=0$; Fig.~\ref{Fig5}a and~b).

Somewhat surprisingly, experimental results show that the orientation tuning of surround suppression in visual cortex is not locked to the preferred orientation ($\theta_0$) of the neuron under examination\cite{Shushruth2012}. If a non-optimal stimulus ($\theta_n$) is presented in the cRF, then the strongest suppression occurs when the grating orientation presented in the surround matches the non-optimal stimuls ($\theta_n$), rather than the neuron's preferred orientation ($\theta_0$). This phenomena results in a progressive shift of the surround suppression response tuning curve, such that the minimum of the curve is aligned with the orientation of the stimulus presented in the cRF.

Our model reproduces both these aspects of surround suppression, by combining local competition with orientation-tuned long-range excitatory connections (see Fig.~\ref{Fig5}). Note that long-range excitatory connections are made with no inhibitory bias in our model. That is, synapses are made onto excitatory and inhibitory targets in proportion to the existence of those targets in the cortex, as observed experimentally\cite{Kisvarday_etal86}. In our model, 80\% of local and long-range excitatory synapses are made onto excitatory targets (see Methods). Excitatory synapses are evenly split between local and long-range projections (i.e. $P_{IN}, P_I =\SI{50}{\%}$; see Methods); this is also in line with experimental observations\cite{Binzegger_etal04}.

Consistent with experimental findings in columnar visual cortex, the strongest surround suppression occurs in our model when the stimulus orientation presented in the cRF and in the visual surround are aligned (Fig.~\ref{Fig5}a--b). This is because strongest local competition is recruited when the long-range connections from the visual surround are activated simultaneously with local input. Responses in our model are suppressed on average under wide-field stimulation (see Fig.~\ref{Fig5}a). Due to the competitive mechanism responsible for suppression in our model, one of the local subnetworks will have stronger activity than the others. As a consequence, a subset of neurons express facilitation under surround stimulation (suppression index $SI<0$ in Fig.~\ref{Fig5}b). The proportional size of the facilitated population depends on the number of local subnetworks (four in our model, implying that $1/4$ of excitatory neurons exhibit facilitation).

Our model replicates the phenomenon of maximum suppression shifting with the orientation presented to the cRF (Fig.~\ref{Fig5}c; experimental observations in \cite{Shushruth2012}). When non-preferred stimuli are presented in the cRF (pips in Fig.~\ref{Fig5}c), the profile of suppression shifts in response such that strongest suppression occurs when the surround orientation matches the cRF orientation (colored curves in Fig.~\ref{Fig5}c).

\subsection{Local competition explains sparsening of local responses to wide-field natural stimuli.}

Responses of neurons in visual cortex are poorly correlated in response to natural stimuli\cite{Vinje_Gallant00, Yen2007, Haider_etal10, Martin_Schroder13}, and even more negatively correlated in response to wide-field stimulation compared with cRF-only stimulation\cite{Vinje_Gallant00, Haider_etal10}. Reduced correlation of responses leads to increased population sparseness, increasing information coding efficiency of cortex as discussed above. At the same time, lifetime sparseness also increases with wide-field stimulation\cite{Vinje_Gallant00, Haider_etal10} --- this further improves the selectivity of neurons in cortex, by ensuring they fire in response to only few configurations of visual stimuli. It should be noted that population and lifetime sparseness are not necessarily correlated in populations of neurons\cite{willmore2011sparse}, meaning that increases in one do not imply a corresponding increase in the other measure.

\begin{figure}
\centering
\includegraphics[angle = 0, width = 5.0 in]{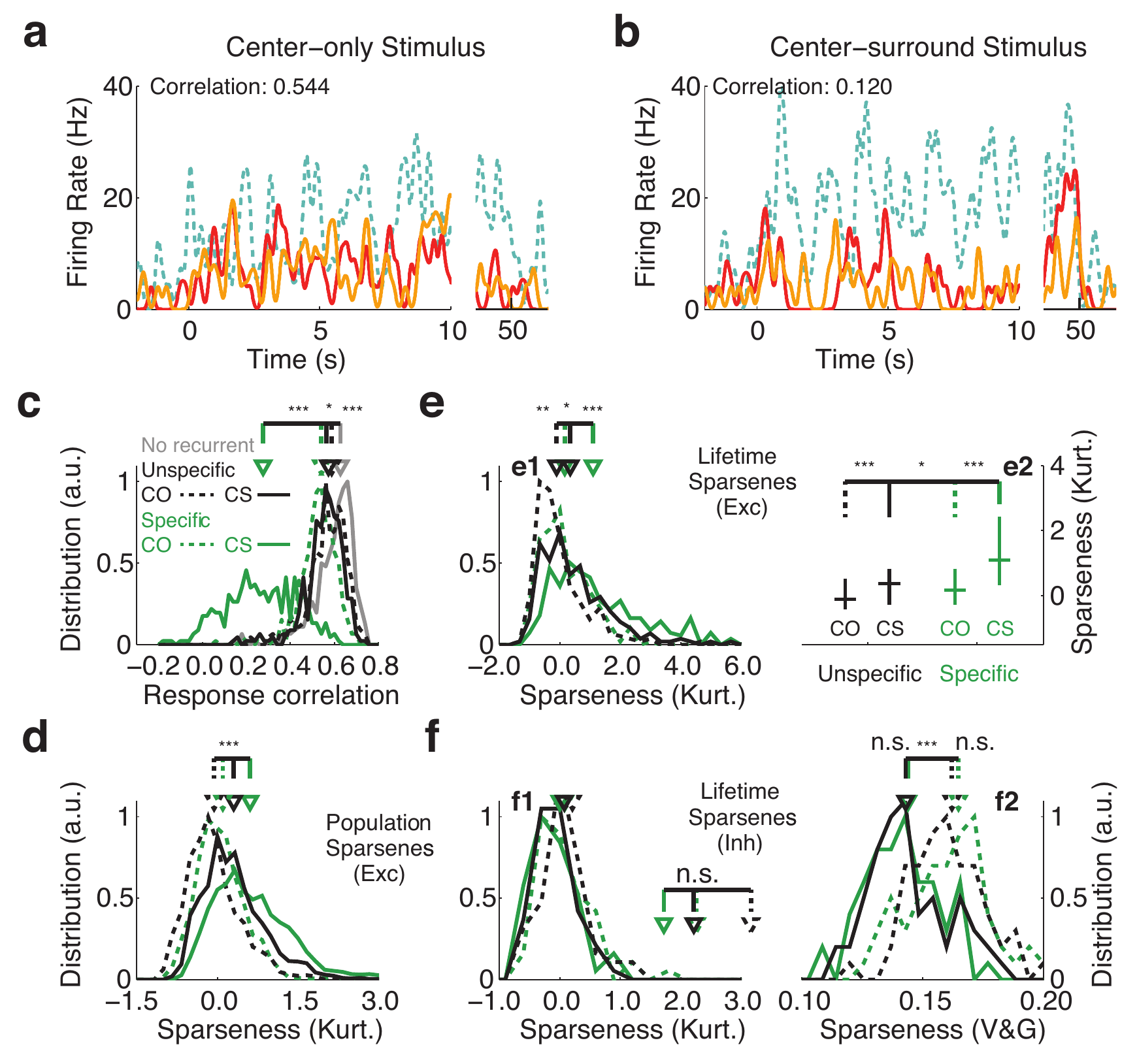}
\caption[Responses of neurons to the time-varying natural stimulus.]{\textbf{Sparsening of responses to simulated natural stimuli.} \textbf{(a--b)}~Firing rate profiles in response to cRF-only~(a) and wide-field natural stimuli~(b). Colors are as indicated in Fig.~\ref{Fig2}a--d. Firing rate curves show the responses of neurons tuned for~\ang{90} orientation, as indicated in a and~b. See Methods for a description of the time-varying natural stimulus. \textbf{(c--e)}~Wide-field stimulation provokes a significant reduction in correlations within the local population~(c), reflected in a significant increase in both population (d) and lifetime sparseness~(e). Colors and curve styles as indicated in (e2). \textbf{(f)}~Responses of the inhibitory population to wide-field stimulation are significantly elevated, and are less sparse under the Vinje-Gallant measure (right;~V\&G) but not under a kurtosis measure (left; Kurt.). Inset: statistical comparison for (f1). Stronger and less sparse inhibitory responses have been observed experimentally\cite{Vinje_Gallant00, Haider_etal10}. ``Specific'': full network. ``Unspecific'': network without local competition in c--f,~$P_+=P_M = \SI{25}{\%}$, with other parameters unchanged. ``No recurrent'': network with all recurrent connections removed,~$J_E=J_I=0$. a.u.: arbitrary units. Horizontal bars indicate significance; n.s.: not significant,~$p>0.05$; *~$p<0.05$; ***~$p<0.001$.
}
\label{Fig6}
\end{figure}

We probed our model with simulated natural stimuli, presented either to the central population only, or as wide-field stimuli (Fig.~\ref{Fig6}). Responses of a column of neurons to center-only stimulation were significantly less correlated than the correlations present in the input stimulus, measured by recording the responses of a control network with no recurrent connectivity (Fig.~\ref{Fig6}c; med. correlation coefficients~$0.59$ vs~$0.63$; $P<10^{-10}$, rank-sum test, 4~neurons $\times$ 60~columns $\times$ 15~trials). However, wide-field stimulation further reduced response correlations within a column (med. correlation~$0.28$; $P\approx0$ versus centre-only stimulation, rank-sum test, 4~neurons $\times$ 60~columns $\times$ 15~trials). This decorrelation led to a significant increase in population sparseness in response to wide-field stimulation (Fig.~\ref{Fig6}d; med. sparseness~$0.11$ vs $0.62$; $P\approx0$, rank-sum test, 240~neurons $\times$ 15 trials), consistent with experimental observations\cite{Vinje_Gallant00, Yen2007}.

The response selectivity of excitatory neurons, measured by lifetime sparseness, was also increased in our model under wide-field stimulation (Fig.~\ref{Fig6}e; med. sparseness $1.10$ vs $0.17$; $P\approx0$, rank-sum test, 240~neurons $\times$ 15~trials), also consistent with experimental observations\cite{Vinje_Gallant00, Yen2007, Haider_etal10}. This increase in excitatory selectivity came at the cost of inhibitory selectivity (Fig.~\ref{Fig6}f). Inhibitory activity was increased on average between cRF and wide-field stimulation (mean response $15.5$ vs~\SI{18.3}{Hz}; $P\approx0$, t-test, 60~neurons), and lifetime sparseness decreased slightly but not significantly under the kurtosis measure and decreased significantly under the Vinje-Gallant measure (med. sparseness $0.14$ vs $0.17$; $P<10^{-7}$, rank-sum test, 60~neurons $\times$ 15~trials; see Methods). This inverse relationship between excitatory and inhibitory selectivity is consistent with experimental observations\cite{Haider_etal10}.

\section*{Discussion}
We constructed a model for columnar visual cortex, that proposes a mechanism for integration of information from the visual surround which is consistent with both recent neuro-anatomical and -physiological measurements. Specifically, the network proposed is consistent with the known meso-scale architecture of columnar cortex, which is characterised by long-range, functionally specific (i.e. orientation specific) lateral excitatory projections coupled with short-range local inhibition\cite{mcguire1991targets, kisvarday1993functional, Haider_etal10, Muir_Cook14, Cossell_etal15}. We include excitatory specificity over a set of local excitatory subnetworks (SSN-specificity) in order to explore the effect of local competition within a cortical column\cite{Muir_Cook14}. 

\subsection{Local competition versus inhibitory specificity.}
Several previous models of surround visual interactions have proposed alternative mechanisms for surround suppression. Schwabe and colleages explained the suppressive effects of far surround visual stimulation through fast axonal transport over inter-areal connections\cite{Schwabe_etal06}, but did not examine orientation tuning of surround suppression. They also require that long-range excitatory projections preferentially target inhibitory neurons, which is not justified by anatomical studies of columnar visual cortex\cite{Somogyi_etal83} but which may underlie a portion of surround suppression in the rodent\cite{Adesnik_etal12}. Shushruth and colleagues proposed a model that reproduces the fine orientation tuning of surround suppression\cite{Shushruth2012}, which relies on strongly tuned feedforward inhibition from the visual surround in a hand-crafted inhibitory network. Several other models for surround suppression (e.g. Somers and colleagues\cite{Somers_etal98}; Stetter and colleagues\cite{Stetter_etal00}; Rubin and colleagues\cite{rubin2015stabilized}) assume horizontally-expressed inhibition on the spatial scale of orientation hypercolumns -- a region of the cortical surface spanning all preferred orientations, corresponding to around \SI{1}{mm} in cat primary visual cortex. However, competition on this scale is poorly supported by cortical anatomy\cite{Muir_Cook14}.

In our model, suppression provoked by surround stimulation arises through a \textit{local} competitive mechanism, mediated by strong inhibition within the cortical column balancing local and long-range excitation\cite{Binzegger_etal04, Douglas_Martin04}. Competition arises from local SSN-specific excitatory and non-SSN-specific inhibitory recurrent connections, and is recruited by SSN- and $\theta$-specific long-range excitatory to excitatory (E$\rightarrow$E) projections. As a result, the strongest competition --- and as a corollary, the strongest suppression --- is elicited when the center and surround visual fields are stimulated with gratings of the same orientation. Recruitment of local competition via orientation-tuned E$\rightarrow$E connections elicits the shift of suppression with cRF orientation observed experimentally\cite{Shushruth2012}. Importantly, long-range excitatory projections in our model are not class-specific; they target excitatory and inhibitory neurons according to their proportions in the cortex.

\subsection{Sharpening of responses; correlation and information.}
In our model, competition leads to sharpening of response preferences (i.e. increased lifetime sparseness) as well as reducing correlations in activity across the local population (i.e. increased population sparseness), implying that more information about the stimulus is transmitted by each spike. Competition therefore reduces correlation in the sense of signal correlation, which did not occur in a network with local connectivity modified to remove competition (Fig.~\ref{Fig6}c). Inhibitory feedback loops, which are abundant in cortical networks, efficiently reduce correlations in neuronal activity, to the extent that neurons receiving identical presynaptic input can fire nearly independently\cite{Renart_etal10, Tetzlaff2012, Helias2014}.

Using this architecture of specific excitation and non-specific inhibition, our model reproduces many features of visual responses related to surround integration, such as orientation-tuned surround suppression and sparsening of excitatory responses under wide-field stimulation. In addition, local competition between neurons within a column in our model explains the surprisingly low noise correlations between neighbouring neurons in columnar visual cortex\cite{Ecker2010, Martin_Schroder13}.

Natural stimuli are extremely redundant \cite{Kersten87, Schwartz_Simoncelli01}. If responses of neurons in visual cortex reflected this redundancy by simply relaying input stimuli, then the information transmitted per neural spike would be low (i.e. a non-sparse encoding). Instead, responses in visual cortex are decorrelated or anticorrelated, implying the presence of a neuronal or network mechanism that reduces correlations in response to visual stimuli. Some temporal whitening may take place at the level of the retina or LGN\cite{Dan_etal96}, and some low-level reduction of spatial correlation is implemented by surround inhibition in the retina\cite{Hartline49, Kuffler53} and dLGN\cite{Singer_Creutzfeldt70}. However, active correlation reduction of nearby neurons in cortex with similar orientation preference, or between cortical neurons distributed across visual space, cannot occur in either the retina or dLGN. Stimulation of the visual surround leads to increased response sparseness and therefore improved coding efficiency from an information-theory perspective, possibly to reduce the deleterious effect of natural stimulus redundancy\cite{field1987relations}.

\subsection{Competition and response suppression.}
Due to the presence of local competition in our model, the majority of neurons within a column will show suppression at any given time. As a corollary, some minority of neurons will show facilitation in response to surround stimulation (the ``winners'' of the competition). In fact, the proportion of facilitating neurons is directly related to the number of local network partitions that are in competition and to the strength of competition. This result implies that experimental quantification of the proportion of facilitation will provide a direct estimate of that parameter of network connectivity. In our model, only a single subnetwork can win the competition; all other subnetworks are suppressed. Therefore, the proportion of neurons undergoing facilitation will converge to $1/M$, where $1/M$ is an estimate of the size of a local subnetwork and $M$ is an estimate for the number of local subnetworks. This parameter can be estimated directly from in vivo recordings of facilitation and suppression under centre-only and full-field visual stimulation, as long as neurons are roughly evenly distributed between local subnetworks in cortex.

\subsection{Formation of local and long-range specificity.}
Our model assumes that local excitatory neurons form ensembles within which recurrent connections are made more strongly, called specific subnetworks or ``SSNs''. Patterns of local connectivity consistent with a subnetwork architecture have been described in rodent cortex\cite{Yoshimura_Callaway05, Yoshimura_etal05, Ko_etal11, Hofer_etal11, Cossell_etal15}, but have not been examined in animals with columnar visual cortices. Nevertheless, plastic mechanisms within recurrently connected networks of excitatory and inhibitory neurons lead to partitioning of the excitatory network into ensembles\cite{Litwin-Kumar_Doiron14}. This process occurs during development in rodent cortex after the onset of visual experience\cite{Ko_etal13}; we suggest that similar fundamental plastic mechanisms could apply in columnar visual cortex.

Long-range horizontal connections develop in several stages in cat area~17. Initial sparse outgrowth is followed by pruning and increasingly dense arborisation\cite{Callaway_Katz90}, with the result that regions of similar orientation preference are connected\cite{Bosking_etal97, Muir_etal11a}. The mechanism for surround suppression exhibited by our model is strongly expressed when long-range excitatory projections between two distant populations preferentially and reciprocally cluster their synapses within individual SSNs in the two populations (i.e. $P_M>0.5$; see Figs~\ref{Fig3} and~\ref{Fig4}). Note that the precise identity of the two SSNs is not crucial; for simplicity we give them the same index in our model.

We propose that specific targetting of long-range projections could come about through similar mechanisms described above for local partitioning of cortex into SSNs. Initially, long-range projections would be made nonspecifically, across SSNs. However, the tendency for local populations to compete will lead to the activity of individual SSNs to be out of phase with each other. SSNs that happen to be concurrently active in two distant populations will induce reciprocal clustering of long-range projections between the two SSNs in the two populations. Concurrently active SSNs will therefore begin to encourage each other's activity, leading to stronger clustering of long-range projections. 

\begin{center}
    \adfhangingflatleafleft
\end{center}

Competitive mechanisms are known to promote sparse coding\cite{Rozell_etal08, king2013inhibitory}; we showed that the architecture of columnar visual cortex lends itself well to local competition as a fundamental computational mechanism for cortex. Local competition can be recruited by excitatory influences from the visual surround to increase response selectivity; this mechanism explains many features of surround visual stimulation in columnar visual cortex. Local excitatory SSN-specificity, over and above connection preference for similar preferred orientations, has not been sought experimentally in mammals with columnar cortices, although it is is important for shaping visual responses in rodent cortex\cite{Cossell_etal15}. Our results suggest that careful experimental quantification of local circuitry, in functionally-identified neurons, will be important to identify the mechanisms of surround integration in columnar visual cortex.

\begin{methods}

\subsection{The spiking center-surround model.}

All parameters used for the spiking simulations are summarised in Table~\ref{Table1}. 

\paragraph{Neuron model} Excitatory ($E$; exc.) and inhibitory ($I$; inh.) neurons are modeled as leaky integrate-and-fire neurons \cite{Furman_Wang08}, characterized by a resting potential ${V_L} =  \SI{-70}{mV}$, a firing threshold $V_\textrm{th} = \SI{-50}{mV}$ and a reset potential $V_\textrm{reset} = \SI{-55}{mV}$. The subthreshold membrane potential dynamics $V_{E,I}(t)$ evolve under the differential equation
\begin{equation}
C_m\frac{\textrm{d}V(t)}{\textrm{d}t} = -g_L\left[V(t) - V_L\right] - I_{syn}(t) \textrm{,}
\label{eqn:LIF_Model}
\end{equation}
where the membrane capacitance $C_m = \SI{0.5}{nF}$ for excitatory neurons and $C_m = \SI{0.2}{nF}$ for inhibitory neurons; the leak conductance $g_L = \SI{25}{nS}$ for excitatory neurons and $g_L = \SI{20}{nS}$ for inhibitory neurons. After firing, the membrane potential $C_m$ is reset to $V_\textrm{reset}$ and held there for a refractory period of $\tau_\textrm{ref}$ seconds. Refractory periods are~$\tau_\textrm{ref} = \SI{2}{ms}$ for excitatory neurons and~$\tau_\textrm{ref} = \SI{1}{ms}$ for inhibitory neurons. ${I_{\textrm{syn}}}(t)$ denotes the total synaptic input current.

\paragraph{Synaptic interactions} Excitatory postsynaptic currents (EPSCs; $I_E$); inhibitory postsynaptic currents (IPSCs; $I_I$); excitatory input currents arising from external network input ($I_\textrm{Ext}$); and noisy background synaptic inputs ($I_\textrm{Back}$) are modelled as
\begin{align}
    {I_{E}}(t) &= {g_E}\left[V(t) - V_\textrm{rev}^E\right]\sum_s{n_s\cdot{S_{E,s}}(t)} \textrm{,} \label{eq:IsynE2EI_S} \\
    {I_{I}}(t) &= {g_I}\left[V(t) - V_\textrm{rev}^I\right]\sum_s{n_s\cdot{S_{I,s}}(t)} \textrm{,} \label{eq:IsynI2EI_S} \\
    {I_\textrm{Ext}}(t) &= {g_\textrm{Ext}}\left[V(t) - V_\textrm{rev}^E\right]{S_\textrm{Ext}}(t)  \textrm{ and} \\
    {I_\textrm{Back}}(t) &= {g_\textrm{Back}}\left[V(t) - V_\textrm{rev}^E\right]{S_\textrm{Back}}(t) \label{eq:I_star}
\end{align}
respectively, where the sums are over the set of input synaptic gating variables $S_*$; $n_s$ is the number of synapses formed in a particular connection; and the reversal potentials are given by~$V_\textrm{rev}^E = \SI{0}{mV}$ and~$V_\textrm{rev}^I = \SI{-70}{mV}$. The synaptic gating variables $S_*(t)$ evolve in response to an input spike train of spike time occurrences $t_p$, under the dynamics
\begin{equation}
    \frac{\textrm{d}S_*(t)}{\textrm{d}t} =  - \frac{S_*(t)}{\tau_*} + \sum\limits_p {\delta \left(t - {t_p}\right)} \textrm{.}
\end{equation}
Here $\tau_*$ are time constants of synaptic dynamics, and are given by $\left\{\tau_{E}, \tau_I, \tau_\textrm{Ext}, \tau_\textrm{Back}\right\} = \left\{\SI{5}{ms}, \SI{20}{ms}, \SI{2}{ms}, \SI{2}{ms}\right\}$ for excitatory, inhibitory, external and background synapses, respectively. Note that axonal conduction times are not considered, such that network interactions are considered to be instantaneous. Synaptic peak conductances $g_*$ are class- and pathway-specific, and are defined in more detail below.

\paragraph{Network architecture} The centre-surround model network consists of several populations~$i \in \left[1\dots{N}\right]$, each representing a hypercolumn of cat primary visual cortex (area 17; see Fig.~\ref{Fig1}). For the sake of simplicity we choose only seven populations in this work, i.e.~$N=7$. As such, each population~$i$ contains several columns, with each column~$k$ corresponding to a preferred orientation~$\theta_k$, where~$\theta_k \in \left[0\dots\ang{180}\right]$ and~$k \in \left[1\dots{N_\textrm{Col}}\right]$, such that~${\theta _k} = \left(k - 1\right)180/{N_\textrm{Col}}$. We take~$N_\textrm{Col}=60$ in this paper. In addition, excitatory neurons within each column are partitioned into a set of~$M$ subnetworks indexed with~$j \in \left[1\dots{M}\right]$. For simplicity, a single excitatory neuron is defined per subnetwork, such that no additional index is needed to distinguish neurons within the same subnetwork. Only a single inhibitory neuron is present in each column~$\theta_k$.

Within a single population $i$, synaptic connections are modulated by similarity of preferred orientation $\Delta\theta$, which has a one-to-one mapping with physical cortical space under the transformation $\ang{180} \approx \SI{1}{mm}$\cite{Muir_etal11a}. To avoid edge effects in our model we adopt a circular topology of preferred orientations $\theta$, with a periodicity of \ang{180}. $\Delta \theta$ is therefore defined as the minimum distance around a ring according to the relative orientation preference of two neurons; that is,~$\Delta\theta = \min (|{\theta _1} - {\theta _2}|,180 - |{\theta _1} - {\theta _2}|)$.

Both local (within-population) and long-range (between-population) excitatory connections are modulated by similarity in orientation preference between source and target neurons ($\Delta\theta$), as well as by whether or not the source and target neurons share subnetwork membership $j$. For orientation-specific connectivity, we use a Gaussian function over $\Delta\theta$, given by
\begin{equation}
\mathcal{G}(\Delta \theta, \sigma_\theta) = \frac{180}{\sqrt{2\pi} \cdot {N_\textrm{Col}} \cdot \sigma_\theta}{\exp \left(\frac{-\Delta\theta^2}{2\sigma_\theta^2}\right) } \textrm{.}
\label{eqn:Gaussian_Connection}
\end{equation}

Excitatory and inhibitory synapses are distributed over connection pathways under a set of functions $n_*(i_1, i_2, j_1, j_2, \Delta\theta)$, which define the number of synapses made between any two neurons. These functions for recurrent excitatory connectivity are defined as
\begin{align}
    n_{E\to{E}}(i_1=i_2, j_2=j_2, \Delta\theta) &= P_\textrm{IN}\cdot{P_+}\cdot{f_E}\cdot{N}_\textrm{Syn,E}\cdot\mathcal{G}\left(\Delta\theta, \sigma_E^\textrm{Local}\right) \label{eq:local_same_ssn}\\
    n_{E\to{E}}(i_1{\ne}i_2, j_2=j_2, \Delta\theta) &= \left(1-P_\textrm{IN}\right)P_M\cdot{f_E}\cdot{N}_\textrm{Syn,E}\cdot\mathcal{G}\left(\Delta\theta, \sigma_E^\textrm{Long}\right){\Big /}(N-1) \label{eq:long_same_ssn}\\
    n_{E\to{E}}(i_1=i_2, j_2{\ne}j_2, \Delta\theta) &= P_\textrm{IN}\left(1-P_+\right){f_E}\cdot{N}_\textrm{Syn,E}\cdot\mathcal{G}\left(\Delta\theta, \sigma_E^\textrm{Local}\right){\Big /}(M-1) \label{eq:local_diff_ssn}\\
    n_{E\to{E}}(i_1{\ne}i_2, j_2{\ne}j_2, \Delta\theta) &= \left(1-P_\textrm{IN}\right)\left(1-P_M\right){f_E}\cdot{N}_\textrm{Syn,E}\cdot\mathcal{G}\left(\Delta\theta, \sigma_E^\textrm{Long}\right){\Big /}(N-1)(M-1) \label{eq:long_diff_ssn}
\end{align}
These functions define rules for excitatory connections that are local and within the same SSN (equation~\ref{eq:local_same_ssn}); long-range and within the same SSN (equation~\ref{eq:long_same_ssn}); local across different SSNs (equation~\ref{eq:local_diff_ssn}); and long-range across different SSNs (equation~\ref{eq:long_diff_ssn}) respectively. Connection fields are modulated by parameters~$\sigma_*$; ``Local'' denotes connections within the same population and ``Long'' denotes connections between populations.

Note that a fraction~$P_\textrm{IN}\in\left[0\dots1\right]$ of excitatory synapses are formed locally, and the remainder are distributed across the~$N-1$ other populations. Similarly, fractions~$P_+\in\left[0\dots1\right]$ and~$P_M\in\left[0\dots1\right]$ of excitatory synapses are formed within the same SSN, while the remainder are distributed across the~$M-1$ other SSNs;~$P_+$ controls local subnetwork-specificity, while~$P_M$ controls long-range subnetwork specificity.

Similarly, connectivity functions involving inhibitory neurons are defined as
\begin{align}
    n_{E\to I}(i_1=i_2,\Delta\theta) &= P_I\cdot{f_I}\cdot N_\textrm{Syn,E}\cdot\mathcal{G}\left(\Delta\theta,\sigma_E^\textrm{Local}\right) \label{eq:etoi_local}\\
    n_{E\to I}(i_1\ne i_2,\Delta\theta) &= \left(1-P_I\right)\cdot{f_I}\cdot N_\textrm{Syn,E}\cdot\mathcal{G}\left(\Delta\theta,\sigma_E^\textrm{Long}\right)/(N-1) \label{eq:etoi_long}\\
    n_{I\to E}(\Delta\theta) &= f_E\cdot N_\textrm{Syn,I}\cdot\mathcal{G}\left(\Delta\theta,\sigma_I\right)/M \label{eq:itoe}\\
    n_{I\to I}(\Delta\theta) &= f_I\cdot N_\textrm{Syn,I}\cdot\mathcal{G}\left(\Delta\theta,\sigma_I\right) \label{eq:itoi}
\end{align}
These functions define~$E\to I$ connections made within (equation~\ref{eq:etoi_local}) and between (equation~\ref{eq:etoi_long}) populations;~$I\to E$ connections (equation~\ref{eq:itoe}); and recurrent inhibitory connections (equation~\ref{eq:itoi}). Note that~$P_I\in\left[0\dots1\right]$ excitatory synapses are made with local inhibitory targets, with the remainder made with long-range inhibitory targets. In addition, connections involving inhibitory neurons are~$\theta$-specific but not subnetwork-specific, and inhibitory projections are made only locally (i.e. within the same population).

We therefore define~${I_{\textrm{Syn},E}}(i,j,{\theta_k},t)$ as the total synaptic current input to the excitatory neuron in population $i$, subnetwork~$j$, with preferred orientation~$\theta_k$, at time~$t$. Similarly, we define~${I_{\textrm{Syn},I}}(i,{\theta_k},t)$ as the total synaptic current input to the inhibitory neuron in population~$i$, in the column with preferred orientation~$\theta_k$. Input currents~$I_{\textrm{Syn},*}$ are of the form
\begin{equation}
    I_{\textrm{Syn},*} = I^{\textrm{Local}}_* + I^{\textrm{Long}}_* + I_{\textrm{Ext}\to{*}} + I_{\textrm{Back}\to{*}}\textrm{,} \label{eq:general_currents}
\end{equation}
where  ``Ext'' denotes external inputs provides as stimuli to the network; and ``Back'' denotes background inputs representing spontaneous activity. Each term in equation~\ref{eq:general_currents} evolves according to the synaptic dynamics in equation~\ref{eq:I_star}, weighting input from the rest of the network according to the connectivity functions Eqs~\ref{eq:local_same_ssn}--\ref{eq:itoi}.

\begin{table} \footnotesize
\centering 
\caption[Parameters]{Parameters used in simulations of the spiking model. Exc., $E$: excitatory / excitation; Inh., $I$: inhibitory / inhibition; Prop.: proportion of; Syn.: synapses; SSN: Specific subnetwork.} 
\fontsize{9}{11}\selectfont
\begin{tabular}{| p{14ex} | p{40ex} | p{19ex} |}
\hline  
\textbf{Parameter} & \textbf{Description} & \textbf{Value} \\

\thickhline

$V_L$ & Neuron resting potential & \SI{-70}{mV}\\ \hline
$V_\textrm{th}$ & Neuron firing threshold voltage & \SI{-50}{mV}\\ \hline
$V_\textrm{reset}$ & Neuron reset voltage & \SI{-55}{mV}\\ \hline
$C_{m,E}$; $C_{m,I}$ & Exc. and Inh. neuron membrane capacitance & \SI{0.5}{nF}; \SI{0.2}{nF}\\ \hline
$g_{L,E}$; $g_{L,I}$ & Exc. and Inh. neuron leak conductance & \SI{25}{nS}; \SI{20}{nS}\\ \hline
$\tau_{\textrm{ref}, E}$; $\tau_{\textrm{ref}, I}$ & Exc. and Inh. neuron refractory periods & \SI{2}{ms}; \SI{1}{ms}\\

\thickhline

$g_{E\to E}$; $g_{E\to I}$; $g_I$ & $E\to E$, $E\to I$ and Inh. syn. conductances & \SI{0.05}{nS}; \SI{0.2}{nS}; \SI{0.12}{nS}\\ \hline
$g_\textrm{Ext}$ & Synaptic conductance for external inputs & \SI{11.43}{nS}\\ \hline
$g_\textrm{Back}$ & Synaptic conductance for spontaneous inputs & \SI{11.43}{nS}\\ \hline

\thickhline

$V_\textrm{rev}^E$; $V_\textrm{rev}^I$ & Exc. and Inh. syn. reversal potentials & \SI{0}{mV}; \SI{-70}{mV}\\ \hline
$\tau_{E}$; $\tau_I$; $\tau_\textrm{Ext}$; $\tau_\textrm{Back}$ & Time constants governing syn. dynamics & \SI{5}{ms}; \SI{20}{ms}; \SI{2}{ms}; \SI{2}{ms}\\

\thickhline

$N$ & Number of populations (hypercolumns) & 7\\ \hline
$N_\textrm{Col}$ & Number of columns in each population & 60\\ \hline
$M$ & Number of subnetworks (SSNs) & 4\\ \hline

\thickhline
$f_{E}$ & Prop. syn. made by all neurons onto Exc. targets & $80\%$ \\ \hline
$f_{I}$ & Prop. syn. made by all neurons onto Inh. targets & $20\%$ \\ \hline
$P_{IN}$ & Prop. E$\rightarrow$E syn. made locally & $50\%$ \\ \hline
$P_+$ & Prop. local E$\rightarrow$E syn. reserved to be made within the same SSN & $95\%$ \\ \hline
$P_M$ & Prop. long-range E$\rightarrow$E syn. reserved to be made within the same SSN & $95\%$  \\ \hline
$P_I$ & Prop. E$\rightarrow$I syn. made locally, versus long-range E$\rightarrow$I projections & $50\%$ \\

\thickhline

$\sigma_E^\textrm{Local}$; $\sigma_E^\textrm{Long}$; $\sigma_I$ & Functional tuning of orientation-specific connections & \ang{20}; \ang{20}; \ang{20} \\ 

\thickhline

$N_{\textrm{Syn},E}$; $N_{\textrm{Syn},I}$ & Total syn. made by each exc. or inh. neuron & 3000; 4500 \\

\hline

\end{tabular}
\label{Table1} 
\end{table}

\paragraph{Background noise and stimulation protocol}
All neurons receive an excitatory background noise input, modeled as independent Poisson spike trains at a rate of $v_{\textrm{Back} \to E} = \SI{180}{Hz}$ for excitatory neurons and $v_{\textrm{Back} \to I} = \SI{50}{Hz}$ for inhibitory neurons.

Our model of cat primary visual cortex (area 17) is designed to explore the computational effects of long-range synaptic input from the visual surround on local representations of orientation preference in the superficial layers of cortex. We therefore assume that orientation preference itself is computed within layer 4, and that inputs to the neurons in the superficial layers are already tuned for orientation.

We simulated external visual stimuli as independent Poison spike trains. For oriented grating stimuli, the rate $v_\textrm{Grat}({\theta _k},t)$ of the input spike trains received by both excitatory and inhibitory neurons depended on the preferred orientation $\theta_k$ of the neuron and the instantaneous stimulus orientation $\theta_\textrm{Grat}(t)$, under
\begin{equation}
v_\textrm{Grat}({\theta _k},t) = \alpha _\textrm{Grat}\cdot h(t)\exp \left[ - \frac{{{{\Delta\theta({\theta _k}, {\theta _\textrm{Grat}(t)})}^2}}}{{\sigma _\textrm{Grat}^2}}\right] \textrm{,}
\end{equation}
where~$\sigma_\textrm{Grat} = \ang{27}$ is the tuning sharpness of orientation-tuned inputs, and~$\Delta\theta$ describes orientation differences around the ring topology as described above. The amplitude of stimulus input~$\alpha _\textrm{Grat} = \SI{270}{Hz}$ and~\SI{29}{Hz} for excitatory neurons and inhibitory neurons, respectively. During center-only visual stimulation,~$h(t)=1$ for the center population and~$0$ for surround populations. During center-surround stimulation,~$h(t)=1$ for both center and surround populations. 

The simulated natural stimulus was generated as a complex pattern of varying oriented input over the visual field, which shifted over time. Neurons within the same column received similar input $v_\textrm{Nat}(\theta_k,t)$, by virtue of their shared orientation preference, depending on the difference between the orientation of the stimulus $\theta_{\textrm{Nat},k}$ and their preferred orientation $\theta_k$, under
\begin{equation}
v_\textrm{Nat}({\theta _k},t) = h(t)\left\{{v_\textrm{const}} + {\alpha _\textrm{Nat}}\exp \left[ - \frac{{{{\Delta\theta({\theta _k}, {\theta _{\textrm{Nat},k}(t)})}^2}}}{{\sigma _\textrm{Nat}^2}}\right]\right\} \textrm{,}
\end{equation}
where $\sigma_{Nat} = \ang{20}$. In columnar visual cortex, neighbouring neurons are likely to receive less correlated input than this, due to relative shifts in receptive field location. However since the goal of our model is to investigate local competition and reduction of correlations, we designed our stimulus to contain local correlations. During center-only visual stimulation (\SIrange{0}{50}{s}),~$h(t)=1$ and~$0$ for the center and surround hypercolumns, respectively. During center-surround stimulation~$h(t)=1$ for both hypercolumns.~$\alpha_{Nat} = \SI{40}{Hz}$ and~\SI{0}{Hz} for excitatory neurons and inhibitory neurons, respectively, while~$v_\textrm{const} = \SI{148}{Hz}$ and~\SI{29}{Hz} for excitatory neurons and inhibitory neurons, respectively. 

The stimuli provided to each column $\theta_{\textrm{Nat},k}(t)$ for both center and surround populations were generated by spatially and temporally filtering independent white noise signals for each column. An ergodic noise process $\vartheta_k(t)$ was generated for each column, and evolved under the dynamics
\begin{align}
    \vartheta_k(t + \Delta t) &= \vartheta_k(t) + \lambda_\textrm{noise} \eta_k(t) \textrm{ and}\\
    \tau_\textrm{noise} \frac{\textrm{d}{\eta_k(t)}}{\textrm{d}t} &= -\eta_k(t) + \sigma_\textrm{noise}\xi (t)\sqrt {\tau_\textrm{noise}} \textrm{,}
\end{align}
where $\tau_\textrm{noise} = \SI{2}{ms}$, $\sigma_{noise} = \ang{0.18}$ and $\xi(t)$ is a gaussian white noise process with zero mean and unit variance. The time step of the stimulation $\Delta t = \SI{0.1}{ms}$; and $\lambda_\textrm{noise} = 20$. These ergodic noise processes were then spatially filtered under the relationships
\begin{align}
    \theta _{\textrm{Nat},k}(t) &= \arctan \left\{\frac{{\sum\limits_{l = 1}^{N_{Col}} {\sin \left[{\vartheta _l}(t)\right]G\left[{\theta _l}(t),{\theta _k}(t)\right]}}}{{\sum\limits_{l = 1}^{{N_{Col}}} {\cos \left[{\vartheta _l}(t)\right]G\left[{\theta _l}(t),{\theta _k}(t)\right]} }}\right\}
\end{align}
where $G({\theta _l},{\theta _k}) = {A_0}\cdot\exp \left[ -\Delta\theta(\theta_l,\theta_k)^2 / 2 \sigma_\theta^2 \right]$, $A_0 = 0.6$ and $\sigma_\theta = \ang{2}$.

\paragraph*{Parameters for the spiking model}
Axons of pyramidal neurons in cat visual cortex make a roughly even split of boutons between local and long-range arbors\cite{Binzegger_etal04}, providing estimates for $P_{IN}=0.5$ and $P_I=0.5$. Excitatory neurons make $N_{\textrm{Syn},E} = 3000$ total synapses per excitatory neuron; inhibitory neurons make  $N_{\textrm{Syn},I} = 4500$ total synapses per inhibitory neuron\cite{Beaulieu_Colonnier83, Binzegger_etal04}. All neurons in the model connect to excitatory and inhibitory targets roughly in proportion to the prevalence of excitatory and inhibitory neurons in cortex\cite{Kisvarday_etal86}: $f_E=\SI{80}{\%}$ of synapses in the model are made with excitatory targets and $f_I=\SI{20}{\%}$ of synapses are made with inhibitory targets.

Specificity of connections between excitatory neurons has been observed in several cortical areas in the rodent\cite{Yoshimura_etal05, Kampa_etal06, Perin_etal11}, and is correlated with similarity in visual feature preference in rodent visual cortex\cite{Ko_etal11, Hofer_etal11, Cossell_etal15}. The presence of functionally specific connectivity of the type proposed in this paper has not been investigated in columnar visual cortex (e.g. in cat or monkey), leading us to explore a range of specificity levels $P_{+}$ and $P_M$.  We used nominal values of $P_{+} = P_M = \SI{95}{\%}$.

Since inhibitory responses are similarly tuned as excitatory responses in cat primary visual cortex \cite{anderson2000orientation}, all ${\sigma _\theta }$ are \ang{20}. Since the mapping from orientation to physical distance in the central visual field of cat area 17 is approximately \SI{1}{mm} per \ang{180} hypercolumn, this corresponds to a local anatomical projection field of approximately \SI{450}{\micro\meter} width\cite{Muir_etal11a}.

\subsection{Nonlinear dynamical analysis of the system stability and steady-state response.}

Parameters for the mean-field non-spiking model are give in Table~\ref{tab:mean-field-params}.

\paragraph*{Mean-field dynamics}
We used mean-field analysis methods to investigate the dynamics of the center-surround model. First, we introduce the activation function for excitatory and inhibitory nodes, defined as \cite{wong2006recurrent}
\begin{align}
    r \left[I_\textrm{syn}(t)\right] &= \frac{\phi \left[I_\textrm{Syn}(t)\right]}{1 + \tau _{\textrm{ref}}\cdot\phi \left[I_\textrm{Syn}(t)\right]} \textrm{ and} \\
    \phi [I_\textrm{Syn}(t)] &= \frac{{\gamma \cdot I_{\textrm{Syn}}(t) - I_T}}{{1 - \exp \left\{-c\left[\gamma \cdot I_\textrm{Syn} (t) - I_T\right]\right\}}} \textrm{,}
\label{eqn:firing_rate_model}
\end{align}
where $r\left[I_\textrm{Syn}(t)\right]$ is the firing rate of a neuron in response to the instantaneous synaptic input current $I_\textrm{Syn}(t)$; $c$ and $\gamma$ are the curvature and gain factors of the activation function, respectively. The activation function becomes a linear threshold function with $I_T/\gamma$ as the threshold current when $c$ is large. $\tau_\textrm{ref}$ is the refractory period of the neuron, which also determines its maximum firing rate.

We redefine the synaptic gating variables $S_*(t)$ for excitatory and inhibitory synapses in the mean-field model as\cite{wong2006recurrent}
\begin{gather}
    \frac{\textrm{d}S_*(t)}{\textrm{d}t} = -\frac{S_*(t)}{\tau_*} + r\left[{I_\textrm{*,Syn}}(t)+I_\textrm{*,Ext}+I_\textrm{*,Back}\right] \textrm{,}\label{eqn:DifferentialEquationS}
\end{gather}
in which the total synaptic input currents $I_\textrm{*,Syn}(t)$ are given by
\begin{align}
    I_*(t) &= J_*\sum_s{n_s\cdot S_{*,s}(t)} \textrm{, where} \\
    J_* &= g_*\left[\langle V(t)\rangle - V_\textrm{rev}^*\right] \textrm{,}
\end{align}
and we assume the average membrane potential for each neuron~$\langle V(t)\rangle \approx\SI{-52.5}{mV}$. The parameters $J_E$ and $J_I$ therefore represent constant synaptic weights (for excitation and inhibition, respectively), rather than synaptic conductances as in the spiking model.

\paragraph*{Reduced model}
For simplicity, we present only the analysis of a reduced model, such that we consider only two populations ($N=2$), each containing only two SSNs ($M=2$) and without considering orientation such that ($N_\textrm{Col}=1$; see Fig.~\ref{Fig1}c). The connectivity functions $n_*$ from the spiking model apply as before, but neglecting the indices for preferred orientation. $S_*(t)$ therefore has the form~$S_*(t)=\left[x_{1,1},~ x_{1,2},~ x_{2,1},~ x_{2,2},~ y_1,~ y_2\right]^T\!(t)$. Here~$x_{i,j}(t)$ is the instantaneous value of the gating variable for the excitatory neuron in population~$i$, subnetwork~$j$, and~$y_i(t)$ is the value for the inhibitory neuron in population~$i$.

To investigate the dynamics of the mean-field model we calculate the steady-state responses~$\bar{S}$ of the system, that is~$\bar{S} = S_*(t):\textrm{d}S_*(t) / \textrm{d}t=0$. We solve the simplified system in equation~\ref{eqn:DifferentialEquationS}, under centre-only (``CO'') or centre-surround (``CS'') stimulation, defined as
\begin{align}
    I_\textrm{Ext,CO} &= \left[\iota_E,0,\iota_E,0,\iota_I,0\right]^T \textrm{ and} \\
    I_\textrm{Ext,CS} &= \left[\iota_E, \iota_E, \iota_E, \iota_E, \iota_I, \iota_I\right]^T \textrm{,}
\end{align}
where~$\iota_E$ and~$\iota_I$ are stimulation currents delivered to exc. and inh. neurons respectively, to obtain~$\bar S_\textrm{CO}$ and~$\bar S_\textrm{CS}$. We then numerically obtain the system Jacobian~$J_{\bar S_*}$ at these fixed points, and examine the eigenvalues of~$J_{\bar S_*}$ to determine the stability of the system around these fixed points.

\paragraph*{Competition index and computational regimes}
We also define a competition index (``CI'') to quantify the strength of competition exhibited between excitatory neurons in different subnetworks. The competition index measures how strongly the activity of an excitatory neuron in one SSN in the ``centre'' population is suppressed, when input to the other subnetwork is increased, either for centre-only or centre-surround stimulation. This index is defined as
\begin{equation}
    CI \equiv \frac{\textrm{d}{\bar S}_{1,1}}{\textrm{d}\Delta I_C}
\end{equation}
where $\Delta I_C$ defines a perturbation in the input to a selected set of neurons in the network. In the case of centre-only stimulation, $\Delta I_C = \left[0, 1, 0, 0, 0, 0\right]^T$; $CI$ therefore quantifies the suppression evoked in the neuron in population~1, SSN~1 by an increase in input to SSN~2 in population~1. That is, population~1 is defined as the ``central'' population, and we measure competition between SSNs within that population.

In the case of centre-surround stimulation, $\Delta I_C = \left[0, 1, 0, 1, 0, 0\right]^T$; $CI$~therefore quantifies the suppression provoked by input to the SSN~2 in both populations~1 and~2. The competition index~$CI$ is only defined in the case of stable~$\bar{S}$, i.e. when the eigenvalues of~$J_{\bar S}$ have non-positive real parts.

We use the eigenvalues of $J_{\bar S_*}$ in conjunction with the $CI$ to identify parameter regimes of stability and computation (see Fig.~\ref{Fig4}). If $J_{\bar S_*}$ has one or more eigenvalues with positive real part, the system operates in a hard winner-take-all regime (``hWTA''), in which only a single SSN is permitted to be simultaneously active. This is because competition between SSNs is so strong, that activity of a single SSN is capable of entirely suppressing the activity of the other SSNs, via shared inhibitory feedback.

Alternatively, if all eigenvalues of $J_{\bar S_*}$ are negative, then the system operates in either a ``soft'' winner-take-all regime (``sWTA''; if $CI<0$), such that several simultaneously active SSNs are permitted at steady state, or in a non-competitive regime (``NC''; if $CI>0$). In the sWTA regime, increasing the external stimulus to one SSN will lead to the decreasing of neural activities of the remaining SSNs, implying competition exists between neurons within a column. Stronger competition is indicated by more negative $CI$. In the NC regime, however, increasing the input to one SSN will increase the activity of the remaining SSNs. The absence of competition is reflected in a positive $CI$.

When strong excitation is unbalanced by inhibition, the network is in an ``unbalanced'' regime (``UN''). This regime is defined as when firing rates of all neurons are close to saturation in the steady state, and all eigenvalues are negative.

\paragraph{Parameters for the firing-rate model} We estimated parameters values from experimental measurements of the properties of cortical neurons. The slope of the I--F curve of an adapted cortical pyramidal neuron, corresponding to~$\gamma_E$ in~Eqn.~\ref{eqn:firing_rate_model} is approximately \SI{66}{\hertz\per{\nano\ampere}}\cite{Ahmed_etal98}.  The corresponding value for basket cells~($\gamma_I$) is approximately \SI{351}{\hertz\per{\nano\ampere}}\cite{Nowak_etal03}.  The I--F curvature parameters~$c_E$ and~$c_I$ were chosen to approximate the spiking model. The strength of a single excitatory synapse is estimated by the charge injected into a post-synaptic neuron by a single spike, given by~$I_\textrm{Syn} = J_E\cdot S_E(t)$. At steady-state,~$S_E(t) = \tau_E$, therefore $J_E = I_\textrm{Syn}/\tau_E$. We estimated nominal values of~$J_E = \SI{2.6}{pA}$ and~$J_I = \SI{2.1}{pA}$\cite{Binzegger_etal09}.  We estimated firing thresholds for our neurons of $I_{T,E}/\gamma_E = \SI{0.4}{nA}$ and $I_{T,I}/\gamma_I = \SI{0.2}{nA}$\cite{Ahmed_etal98, Nowak_etal03}.  The average input currents~$\iota_E$ and~$\iota_I$ injected during visual input were estimated from the average currents received by single pyramidal neurons in visual cortex during visual stimulation\cite{Ahmed_etal98}.  We used values of~$\iota_E = \SI{0.90}{nA}$, and a proportional value of~$\iota_I = \SI{0.087}{nA}$\cite{Ahmed_etal98, Binzegger_etal04}.

\begin{table} \footnotesize
\centering 
\caption[Parameters]{Parameters used in analysis of the mean-field model that differ from those given in Table~\ref{Table1}.} 
\fontsize{10}{12}\selectfont
\begin{tabular}{| p{13ex} | p{40ex} | p{14ex} |}
\hline  
\textbf{Parameter} & \textbf{Description} & \textbf{Nominal value} \\

\thickhline

$c_E$; $c_I$ & Exc. and Inh. neuron activation function curvature parameters & \SI{160e-3}{}; \SI{87e-3}{} \\ \hline
$I_{T,E}/\gamma_E$; $I_{T,I}/\gamma_I$ & Exc. and Inh. threshold currents &  \SI{0.4}{nA}; \SI{0.2}{nA} \\

\thickhline

$\langle V(t)\rangle$ & Assumed average membrane potential & \SI{-52.5}{mV} \\ \hline
$J_E$; $J_I$ & Exc. and Inh. total synaptic weights & \SI{2.6}{pA}; \SI{2.1}{pA} \\

\thickhline

$N$ & Number of populations (hypercolumns) & 2\\ \hline
$N_\textrm{Col}$ & Number of columns in each population & 1\\ \hline
$M$ & Number of subnetworks (SSNs) & 2\\

\thickhline

$\iota_E$; $\iota_I$ & Input currents representing external stimuli to Exc. and Inh. neurons & \SI{0.90}{nA}; \SI{0.087}{nA} \\

\hline

\end{tabular}
\label{tab:mean-field-params} 
\end{table}

\subsection{Population and lifetime sparseness measures.}
The measure of the population sparseness and the life-time sparseness we used mainly is the kurtosis, which measures the~$4^{th}$ moment relative to the variance squared \cite{olshausen2004sparse} and is given by
\begin{equation}
    s_k = \frac{1}{n}\sum\limits_i {\frac{(r_i - \bar r)^4}{\sigma ^4}} - 3 \textrm{,}
\end{equation}
where~$r_i$ is the firing rate of each neuron during the presentation of the~$i^{th}$ natural stimulus, and~$n$ is the number of natural stimuli frames for lifetime spareness. For population sparseness,~$r_i$ is the firing rate of neuron~$i$ during a frame of natural stimuli, and~$n$ is the number of simultaneously recorded neurons in our model.~$\bar r$ is the mean firing rate and~$\sigma$ is the standard deviation of the firing rate. For a sparse (i.e. heavy-tailed) distribution,~$s_k>0$.

In addition we used the Vinje-Gallant measure for sparseness, a nonparametric statistic employed previously in \cite{Vinje_Gallant00, Yen2007}, given by
\begin{equation}
s_\textrm{V\&J} = {\left[ 1 - {\left(\sum\limits_i {\frac{{{r_i}}}{n}} \right)^2} {\Bigg /} {\left(\sum\limits_i {\frac{{r_i^2}}{n}} \right)}\right]} {\Bigg /} {\left(1 - 1/n\right)} \textrm{,}
\end{equation}
where~${s_\textrm{V\&J}} \in [0,1]$. A larger~$s_\textrm{V\&J}$ indicates a more sparse response.

\subsection{Suppression index.}
The strength of surround suppression was quantified using a suppression index ($SI$), which is defined as 
\begin{equation}
    SI\left(\theta\right) = 1 - \frac{R_{CS}\left(\theta\right)}{R_{CO}}\textrm{,}
\end{equation}
where $\theta$ is the center-surround orientation difference, $R_{CO}$ is the response to the center-only stimulus, $R_{CS}(\theta)$ is the response to the center-surround stimulus. Therefore, $SI=1$ indicates that the response in the center population is completely suppressed by the surround stimulus, whereas $SI=0$ indicates the absence of any suppression from the surround.

\subsection{Statistical methods.}

All tests are non-parametric two-sided tests of medians (Wilcoxon Rank Sum), unless stated otherwise.

\end{methods}


\nolinenumbers

\bibliography{natcomms.bbl}


\begin{addendum}
\item The authors would like to thank Elisha Ruesch, Nuno da Costa and Kevan Martin for providing the orientation preference map in Fig.~\ref{Fig1}. The authors would also like to thank the participants of the Capo Caccia workshop (\url{http://capocaccia.ethz.ch}) for many useful discussions that guided this work. This work was supported by the Fundamental Research Funds for the Central Universities (to HY); NSFC (grant 31500863 to HY); the 973 project (grant 2013CB329401 to HY); the European Research council (``Neuromorphic Processors'' neuroP project, grant 257219 to GI); and the CSN (fellowships to DRM). The funders had no role in study design, data collection and analysis, decision to publish, or preparation of the manuscript.
 \item[Competing Interests] The authors declare that they have no
competing financial interests.
\item[Correspondence] Correspondence and requests for materials should be addressed to D.R.M.~(email: dylan.muir@unibas.ch).
\end{addendum}


\end{document}